\documentclass[pdflatex,sn-mathphys]{sn-jnl}
\usepackage{graphicx}
\usepackage{siunitx}
\usepackage{amssymb}
\usepackage{cleveref}
\usepackage{float}
\usepackage[shortlabels]{enumitem}

\newcommand{\showcomment}[2]{{\color{#1}{{#2}}}}
\newcommand{\jz}[1]{\showcomment{black}{#1}}
\newcommand{\jzz}[1]{\showcomment{black}{#1}}

\begin{document}

\title[~]{An Implantable Piezofilm Middle Ear Microphone: Performance in Human Cadaveric Temporal Bones}
\author[1]{\fnm{John Z.} \sur{Zhang}}\equalcont{These authors contributed equally as first authors to this work.}

\author[2]{\fnm{Lukas} \sur{Graf}}\equalcont{These authors contributed equally as first authors to this work.}

\author[2]{\fnm{Annesya} \sur{Banerjee}}

\author[1]{\fnm{Aaron} \sur{Yeiser}}

\author[2]{\fnm{Christopher I.} \sur{McHugh}}

\author[3]{\fnm{Ioannis} \sur{Kymissis}}

\author[1]{\fnm{Jeffrey H.} \sur{Lang}}

\author[3]{\fnm{Elizabeth S.} \sur{Olson}}

\author*[2]{\fnm{Hideko Heidi} \sur{Nakajima}}\email{heidi\_nakajima@meei.harvard.edu}

\affil[1]{\orgname{Massachusetts Institute of Technology}}

\affil[2]{\orgname{Harvard University and Mass Eye and Ear}}

\affil[3]{\orgname{Columbia University}}

\jyear{2023}

\abstract{\textbf{Purpose:} One of the major reasons that totally implantable cochlear microphones are not readily available is the lack of good implantable microphones. An implantable microphone has the potential to provide a range of benefits over external microphones for cochlear implant users including the filtering ability of the outer ear, \jz{cosmetics}, and usability in all situations. This paper presents results from experiments in human cadaveric ears of a piezofilm microphone concept under development as a possible component of a future implantable microphone system for use with cochlear implants. This microphone is referred to here as a drum microphone (DrumMic) that senses the robust and predictable motion of the umbo, the tip of the malleus.

\textbf{Methods:} The performance was measured of five DrumMics inserted in four different human cadaveric temporal bones. Sensitivity, linearity, \jzz{bandwidth}, and equivalent input noise were measured during these experiments using a sound stimulus and measurement setup.
 
\textbf{Results:} The sensitivity of the DrumMics was found to be tightly clustered across different microphones and ears \jzz{despite} \jz{differences in umbo and middle ear anatomy}. The DrumMics were shown to behave linearly across a large dynamic range \jz{(46~dB~SPL to 100~dB~SPL)} across a wide bandwidth (100 Hz to 8 kHz). The equivalent input noise (over a bandwidth of 0.1 – 10 kHz) of the DrumMic and amplifier referenced to the ear canal was measured to be about 54 dB SPL in the temporal bone experiment and estimated to be 46 dB SPL after accounting for the pressure gain of the outer ear.

\textbf{Conclusion:} 
The results demonstrate that the~DrumMic~behaves robustly across ears and fabrication. The equivalent input noise performance \jz{(related to the lowest level of sound measurable) was shown to approach that of commercial hearing aid microphones. To advance this demonstration of the DrumMic concept to a future prototype implantable in humans, work on encapsulation, biocompatibility, connectorization will be required.}}


\maketitle

\section{Introduction}
\jzz{More than one million cochlear implants (CIs) have been implanted worldwide~\cite{zeng2022celebrating}}, changing the lives of many by enabling them to hear and understand speech. Younger patients, and those with less profound hearing loss, are now benefiting from cochlear implants. More patients are also receiving bilateral CIs, increasing their ability to understand speech in noisy environments~\cite{potts2019improving}. When the sensory mechanism has been greatly compromised (yet the auditory nerve is functioning), CIs can \jz{sometimes} allow better hearing than conventional hearing aids.

Despite advances in CI technology, totally implanting all CI-associated hardware internally has not been generally realized. With a totally implantable CI system, wearers could experience multiple advantages: various activities would not be limited, hearing would be possible all the time (for safety and to increase a child’s auditory development), noise such as that from wind would be reduced, and stigmatization due to external hardware would be diminished~\cite{shinners2008implantable}. If designed accordingly, \jz{the system} could use the natural advantages of the outer ear's pressure gain and filtering, improve speech comprehension in noisy environments, and may enable binaural processing~\cite{shaw1974transformation}. Advances in technology---batteries, non-contact charging, low-power electronics, miniaturization of electronics---all work towards the potential of a totally implantable CI system. However, the major barrier is the absence of an implantable microphone that is sensitive, low noise, consistent, reliable, and uses the benefits of the outer ear\jz{~\cite{trudel2022remaining}}.

\jz{Prior work in this direction has been summarized in the related work section of past theses and review articles~\cite{zhang2008acoustic,hake2021piezoelectric,calero2018technical}}. \jzz{Commercial and experimental implantable microphone devices have limitations. For example, Carina~\cite{pulcherio2014carina} is implanted under the skin, which limits the signal-to-noise ratio due to the skin attenuating the external sound recorded, while picking up internal body noise. The Envoy Esteem senses incus body motion with a needle tip. However, because the incus body moves with varying modes with respect to frequency and its surface is smooth, stability and consistency across varying middle ear anatomy for surgical insertion are of concern~\cite{chung2013optimal}. Experimental devices, such as the intracochlear hydrophone concept proposed by the University of Zurich group~\cite{pfiffner2016mems}, have limited bandwidth (500 Hz to 3.5 kHz) and an equivalent input noise of 65 dB SPL at 1 kHz as reported in~\cite{pfiffner2016mems}.}


To overcome many of the challenges of realizing an implantable microphone and to incorporate the acoustic advantages of the outer ear, we have developed \jz{an implantable microphone concept} that senses the motion of the umbo (the tip of the malleus) where it attaches to the center of the tympanic membrane~\cite{cary2022umbo}. An advantage of sensing umbo motion is that the umbo has the largest predominantly one-dimensional motion among other points along the ossicular chain, and it is a good representation of the external sound~\cite{nakajima2005experimental}. Our \jz{umbo microphone}, \jzz{which we refer to as the Drum Microphone (DrumMic),} is constructed of a piezoelectric polymer, polyvinylidene fluoride (PVDF). The design has a sensitive PVDF membrane ``drum'' fixed at the circular perimeter of a metal ring. Within the middle-ear cavity, the PVDF drum membrane presses against the umbo to sense its motion, and the opposite end of the ring sits on top of the cochlear promontory. This design allows for simple surgical implantation. Our analyses of how the DrumMic functions, studied with two types of computational models (analytical and finite element models), agreed with each other. These two modeling results agreed with bench testing results with an artificial umbo, as described in Cary et al. 2022~\cite{cary2022umbo}. \jz{Preliminary tests} in a human cadaveric temporal bone also provided results that were similar~\cite{cary2022umbo}. In this study, we improve the fabrication of the DrumMic \jz{to produce more consistent and reliable sensors through the use of a laser cutter and gluing procedure to adhere the PVDF to the support structure as described in Methods \Cref{sec:methods}}. We also report on the performance of various DrumMics in different cadaveric specimens.

\section{Materials and Methods}
\label{sec:methods}
\subsection{Microphone Design and Construction}
\begin{figure}[hbt!]
    \centering
    \includegraphics[width=1\linewidth]{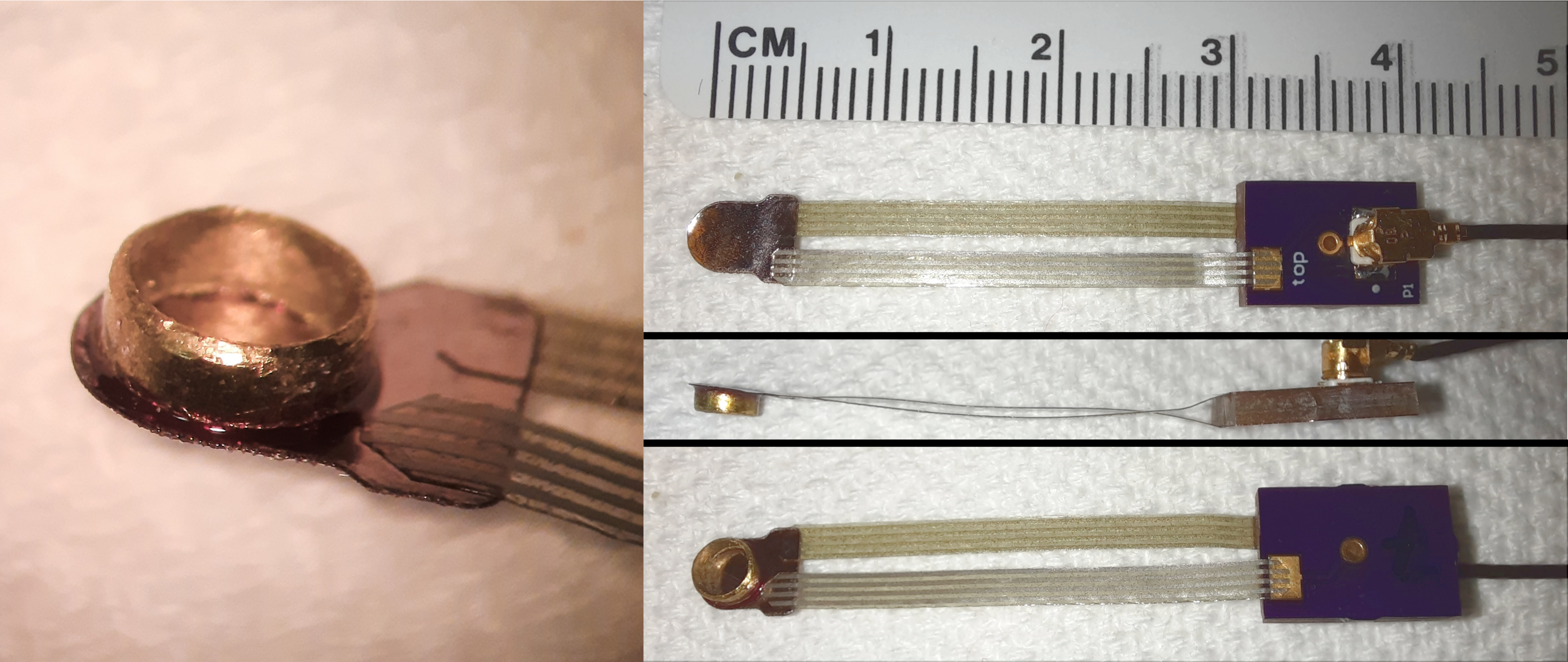}
    \caption{\textit{Left}: View of the DrumMic’s sensitive PVDF membrane with electrodes on both surfaces. A metal ring is glued to one surface of the PVDF. \textit{Right}: Three views of the DrumMic with connectorizations.}
    \label{fig:sensor}
\end{figure}
\jz{We investigate the performance (sensitivity, linearity, noise) of DrumMics subject to variations in fabrication, ear anatomy, and the effect of drumhead static displacement controlled through the number of shims placed \jzz{between the DrumMic base and the cochlear promontory}. Five DrumMics were fabricated and labeled as (1, 2, 3, 4, 5) and experiments were carried out across four different ears labeled (A, B, C, D). We refer to an experiment conducted with a particular ear and DrumMic using a designation such as C4, meaning the combination of ear C and DrumMic 4.} \Cref{fig:sensor} shows our improved design and implementation of the DrumMic. In summary, a PVDF sheet \jz{(PolyK Technologies, Philipsburg, PA, USA)} with copper-nickel electrodes on both sides was laser cut \jz{(Protolaser U4, LPFK, Hannover,
Germany)} into a semicircular shape at one end for the drum, and the other straight end for connectorization. The laser beam ablates a rim of the electrode around the PVDF edge, helping to prevent shorting between the electrode layers. Note that the laser cutter used must have a sufficiently quick pulse duration or the PVDF material will be burned rather than cut. Two of the earliest DrumMics (in experiments A1, B2) were vinyl cut, and were also functional, though vinyl cutting was generally less reliable.

A brass ring of \SI{3}{mm} diameter (with height between \SI{1}{mm} to \SI{1.5}{mm}) was initially held lightly on top of the semicircular-shaped end of the PVDF sheet with a thin layer of cyanoacrylate glue \jz{(Gorilla Glue Company, Cincinnati,
OH, USA)}. Subsequently, the brass ring was firmly fixed in place to the PVDF surface with UV-curing glue \jz{(NEA123 Optical Glue, Norland Products, East Windsor, NJ, USA)} at the outer edge of the brass ring. Care was taken to prevent any adhesive material from seeping onto the PVDF surface inside the brass ring. The PVDF film beyond the edge of the hardened UV-curing glue was trimmed (\SI{0.5}{mm} beyond the outer edge of the brass ring). The resulting dimension of the DrumMic prevented contact to surrounding structures such as the stapes, stapedial tendon, and tensor tympani. On the other end of the PVDF sheet, top and bottom electrodes of the PVDF were connected to flex wire connectors \jz{(HST-9805-210, Elform Inc., Reno, NV, USA)} with a hot bar bonder \jz{(Thin-Line Series 180, HiyachiUnitek, Monrovia, CA, USA)}. The other ends of the flex wires were bonded to a receptacle PCB \jz{(OSHPark, Portland, OR, USA)} that was connected to a custom low-noise charge amplifier~\cite{yeiser2022fully} with a U.FL cable \jz{(Hirose, Tokyo, Japan)}. The flex wires allowed flexibility yet sufficient rigidity during DrumMic insertion to press against the umbo.

To insulate the DrumMic from moisture of the middle-ear cavity, the DrumMic and flex wires were lightly sprayed with silicone \jz{(422B Silicone Modified Conformal Coating Spray, MG Chemicals, Burlington, Ontario, Canada)}. Additionally, the edge of the PVDF with 50-micron separation between the upper and lower electrode layers was brushed with silicone to prevent shorting.

For this experiment we abstained from shielding the DrumMic to electro-magnetic interference (EMI) pickup. While PVDF is FDA approved~\cite{teo2016polymeric}, we disregarded the biocompatibility of certain other materials as the goal of this study was to prove feasibility.

\subsection{Preparation of the Temporal Bone Specimen}
Five fresh (unexposed to fixative) previously frozen human cadaveric temporal bone specimens (2 right, 3 left) were harvested with mean postmortem time of 36.2 hours (standard deviation $\pm$12.2 hours) from 4 male donors (mean age 71.5 years standard deviation $\pm$13.7 years) at Massachusetts General Hospital \jz{(Boston, MA, USA)}. The donors did not have known histories of previous ear pathology, and the ears appeared normal upon inspection.

The middle-ear cavity was accessed through the facial recess in a similar manner used for CI surgery. However, the facial recess was expanded to enable clear views of the middle-ear structures and DrumMic insertion. The middle-ear and inner-ear structures were kept intact.

The bony external auditory canal was shortened to about 1 cm from the posterior tympanic annulus. Two holes were drilled through the bony anterior ear canal (EC) wall: one for a metal tube that connected to the rubber tubing from the speaker, and another for a metal tube sleeve for the microphone probe-tube. Both tubes were glued in place with cyanoacrylate glue. The whole specimen was glued with cyanoacrylate glue to a multi-directional articulated holder \jz{(Noga
Engineering \& Technology, Shlomi, Israel)}.

\subsection{Stimuli and Measurements}
\begin{figure}[hbt!]
    \centering
    \includegraphics[width=1\linewidth]{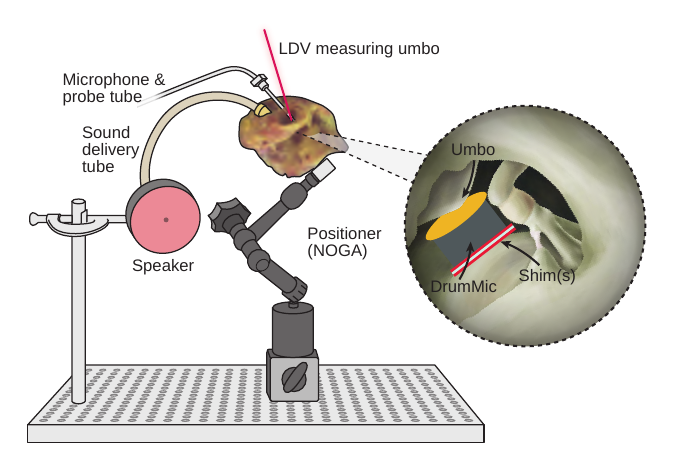}
    \caption{Diagram of the experimental setup. \jz{The cadaveric specimen was held using a NOGA positioner. A speaker was used to produce an ear canal pressure that was measured by a probe-tube microphone. A laser Doppler vibrometer (LDV) was used to measure the motion of the umbo \jzz{(at the center of the tympanic membrane)} through the ear canal.}}
    \label{fig:measurement_setup}
\end{figure}

As shown in \Cref{fig:measurement_setup}, the experimental setup with the specimen was supported by an air table \jz{(TMC, Peabody, MA, USA)} inside a soundproof chamber. The DrumMics were inserted by hand and an adjustable stand was used to provide strain relief for the cables connecting the DrumMic to the amplifier. To prevent drying, moistened Gelfoam pledgets \jz{(Pfizer, New York City, NY, USA)} were placed in the middle ear and mastoid cavity, but not touching middle ear structures or the sensor.

Analog output signals (logarithmic chirps of 0.1--\SI{20}{kHz} or sine sequences of 0.95--\SI{18.2}{kHz}) were produced by a National Instruments DAQ \jz{(National Instruments Austin, TX, USA)}. An audio amplifier \jz{(Crest Audio, Meridian, Mississippi, USA)} drove a vibration-isolated and electrically-shielded Beyer DT48 speaker \jz{(Beyerdynamic, Heilbronn, Germany)} that presented sound to a sealed ear canal.

Ear canal sound near the tympanic membrane was measured with a calibrated Knowles probe-tube microphone (Knowles EK-23103, Itasca, IL, USA), with probe-tube tip approximately 1--3 mm from the umbo. The intensity of the sound stimuli was confirmed to be in the linear range. To measure tympanic membrane motion at the umbo, a laser Doppler vibrometry (LDV) \jz{(CLV 1000, Polytec, Waldbronn, Germany)} beam was aimed at a reflector on the umbo of the tympanic membrane through a clear acrylic cover-slip window that was sealed at its perimeter to the bony ear canal opening. Time and frequency domain measurement signals were collected by a National Instruments DAQ system running custom LabVIEW software \jz{(National Instruments Austin, TX, USA)}.

\subsection{Experimental Protocol}\label{sec:exp_protocol}

\begin{figure}[hbt!]
    \centering
    \includegraphics[width=1\linewidth]{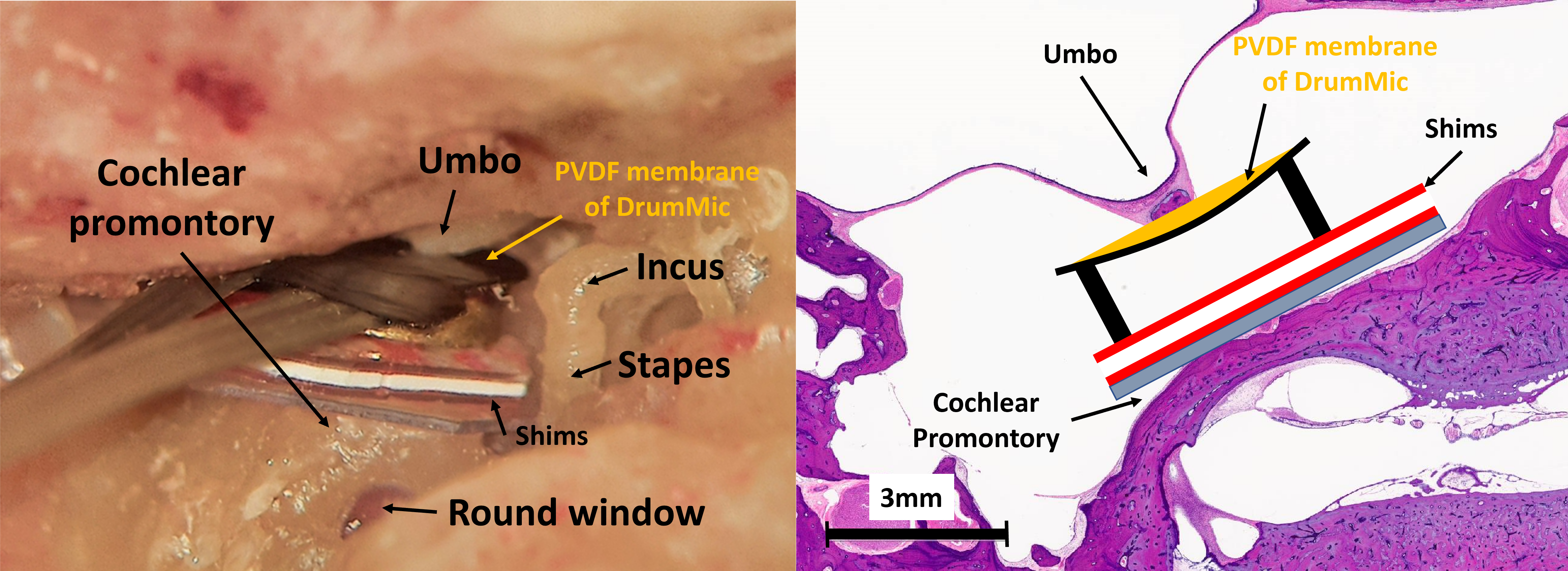}
    \caption{DrumMic implanted in the middle ear. \textit{Left}: View of a left middle-ear cavity with the DrumMic's PVDF membrane interfacing the umbo. Two shims have been inserted between the metal ring base and cochlear promontory. 
    \textit{Right}: Histological cross section with sketch of DrumMic and shims demonstrating how the DrumMic's PVDF membrane interfaces and senses umbo motion. Photo credit: MengYu Zhu for providing the histology picture.}
    \label{fig:histology}
\end{figure}

Initially, to check the integrity of middle-ear-plus-inner-ear input admittance, ear canal umbo velocity referenced to ear canal pressure was confirmed to be normal~\cite{nakajima2005experimental}. In the middle-ear cavity, a DrumMic was inserted between the umbo and cochlear promontory, with the center of the sensitive PVDF DrumMic membrane contacting the umbo, and the open circular bottom of the metal cylinder base sitting on the cochlear promontory (as shown in~\Cref{fig:histology}). Because the distance between the umbo and promontory varied (1.7--\SI{2.3}{mm}~\cite{stieger2006anatomical}), 1--2 plastic shims \jzz{(each shim was 0.65 mm in thickness)} were inserted between the metal cylinder and the promontory. \jz{We investigated the effect of zero, one, or two shims. The number of shims used was limited by the space between cochlear promontory and umbo. The coupling force was not measured, but rather the static displacement of the umbo and the DrumMic was visually observed. Once the deflection was deemed sufficiently large to achieve good sensitivity then no additional shims were needed.} \jz{To insert the shims, first we visually observed that the DrumMic membrane contacted the umbo, then we lifted up the umbo/manubrium to fit the shim(s) under the DrumMic.} The promontory shape and its superior-inferior and lateral-medial position in relation to the umbo varied across specimens. When the promontory was not near symmetric under the umbo, the bottom ring of the DrumMic cylinder had complex motion (e.g., rocking motion). \jz{This} motion could be reduced by gluing the shim to the promontory with carboxylate cement \jz{(Durelon 3M, Saint Paul, MN, USA)}.

The input voltage to the speaker and measurements of ear canal pressure, umbo velocity, and DrumMic (after charge-amplifier) output were simultaneously recorded and averaged 1000 times for chirp \jz{signals} and 20 times for sine-sequences. Noise floor measurements were conducted by recording 10 seconds of voltage output from the DrumMic/charge-amplifier with no acoustic stimuli before and after DrumMic implantation. Control measurements with the speaker tubing decoupled, and sound from the speaker housing \jz{blocked} with putty, verified that electrical coupling was negligible. \jz{Linearity was checked with varying levels of sound input from 35 dB SPL to nearly 117 dB SPL. We demonstrated that the device behaves linearly down to the sensor-amplifier system noise floor so the normalized responses are valid for any input level that still exhibits linearity.}

\subsection{Post-Experiment Data Analysis}
All data was processed using MATLAB (R2022b, MathWorks, Natick, MA, USA). The charge amplifier transfer function was used to convert the measured output voltage of the charge amplifier to charge from the PVDF. A standard calibration method corrected for the effect of the probe-tube to estimate ear canal pressure at the tympanic membrane. LDV velocity data were converted to displacements in the frequency domain. DrumMic input-to-output measurements from several sine sequences and DrumMic harmonics in the frequency spectrum determined linearity. All other frequency response data used log chirp sound stimuli. 

Power spectral densities of the noise were computed from the measured time series data. First, a time series of the amplifier output was recorded by a National Instruments 6122 A/D converter with 16-bit resolution for a full range of $\pm$10 V and sampling rate set to 198 kHz. Next, this was transformed into a power spectral density and then segmented and averaged with a rectangular window following the Bartlett method~\cite{bartlett1948smoothing}. Using the sensitivity transfer function, we converted the output noise-voltage power spectral density to an input-referred noise pressure power spectral density at the tympanic membrane in sound pressure. Equivalent input noise (EIN) spectral densities were integrated numerically to calculate the equivalent input noise in dB SPL. Note that we calculated both the 1/3-octave bandwidth EIN and the EIN across the entire bandwidth from 100 Hz to 10 kHz. By taking into account the outer ear gain, we then computed the EIN referenced to a free field sound illustrating the advantage of implantable microphones that can benefit from the natural filtering of the outer ear. The pressure gain of the outer ear (pinna plus ear canal) was implemented using a polynomial fit to well-known measurements from Shaw ~\cite{shaw1974transformation}.

\subsection{Equivalent Input Noise Calculation}\label{sec:ein}
The equivalent input-referred noise, or simply equivalent input noise (EIN), is commonly used to report noise performance of microphones used in commercial hearing aids such as the Knowles EK-23103 and the Sonion 65GC31. EIN is a useful indicator of the lowest sound pressure level that the microphone can detect.  EIN is derived from a measurement of the noise at the output of the system. It is the acoustic noise presented to a noiseless microphone and amplifier that produces same output as does a noisy microphone and amplifier in the absence of acoustic stimulus. The output-referred noise is then divided by the sensitivity transfer function of the microphone plus amplifier system resulting in the EIN. Finally, the EIN is typically reported in two ways for microphone specifications: (1) as EIN (1/3-octave bands) in a plot over frequency and (2) as a single-number metric EIN (0.1–10 kHz) often A-weighted calculated across the entire bandwidth (see the Brüel \& Kjær handbook~\cite{handbook1995technical}). To compare the EIN of the DrumMic to a realistic hearing aid microphone EIN we present data measured in bench experiments for the Knowles EK-23103, which is a hearing aid microphone that is also the same model of microphone we use as the probe-tube microphone in the temporal bone experiments. In this work, we report the microphone and amplifier system together, though during the design process it is possible to consider each separately.
\section{Results}
\begin{figure}[hbt!]
    \centering
    \includegraphics[width=1\linewidth]{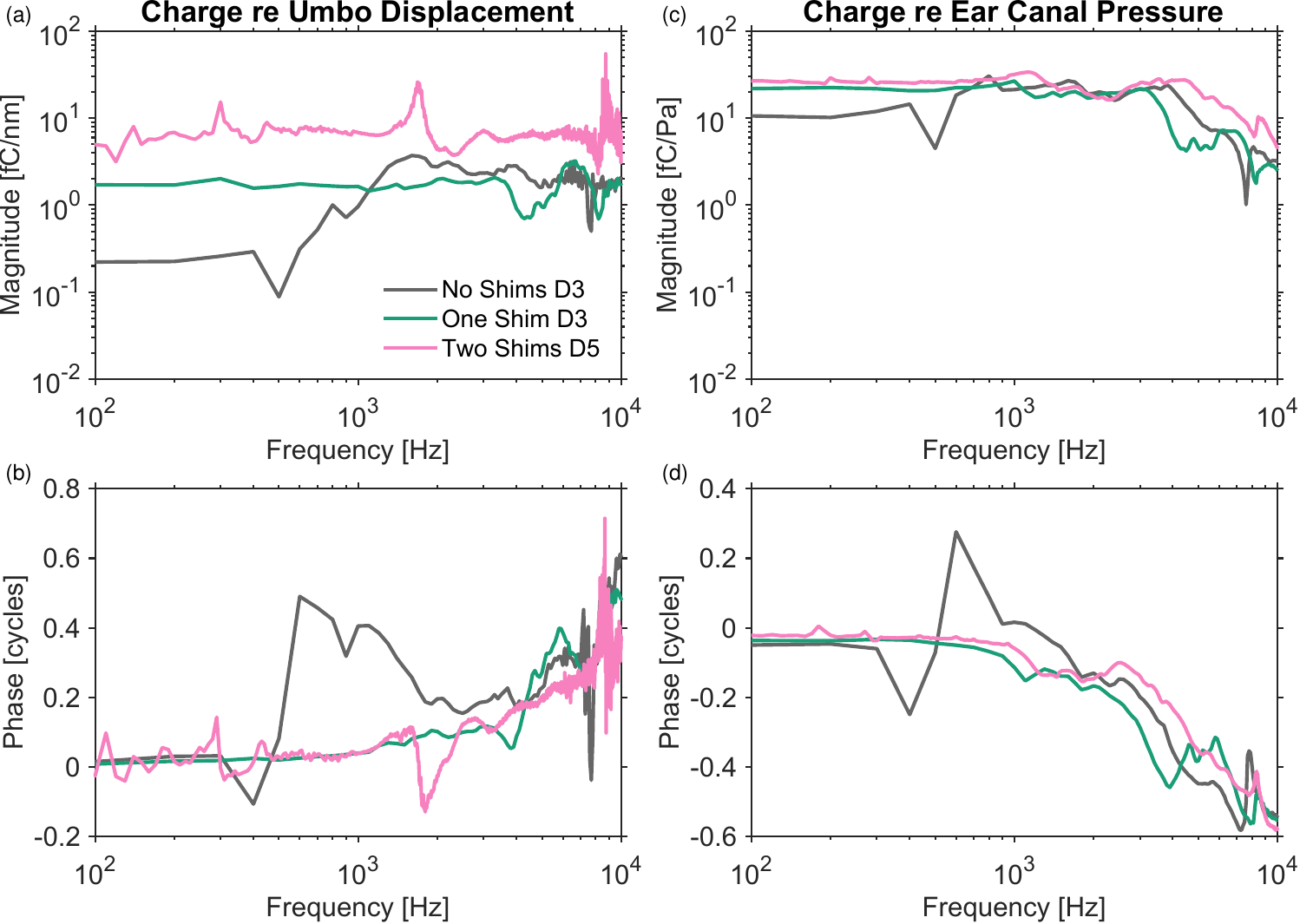}
    \caption{\jz{A representative DrumMic sensitivity with respect to umbo displacement (a) magnitude and (b) phase. DrumMic sensitivity with respect to ear canal pressure (c) magnitude and (d) phase. The output of the DrumMic is electrical charge measured in units of femtocoulombs (fC).}}
    \label{fig:offset}
\end{figure}

\begin{figure}[hbt!]
    \centering
    \includegraphics[width=1\linewidth]{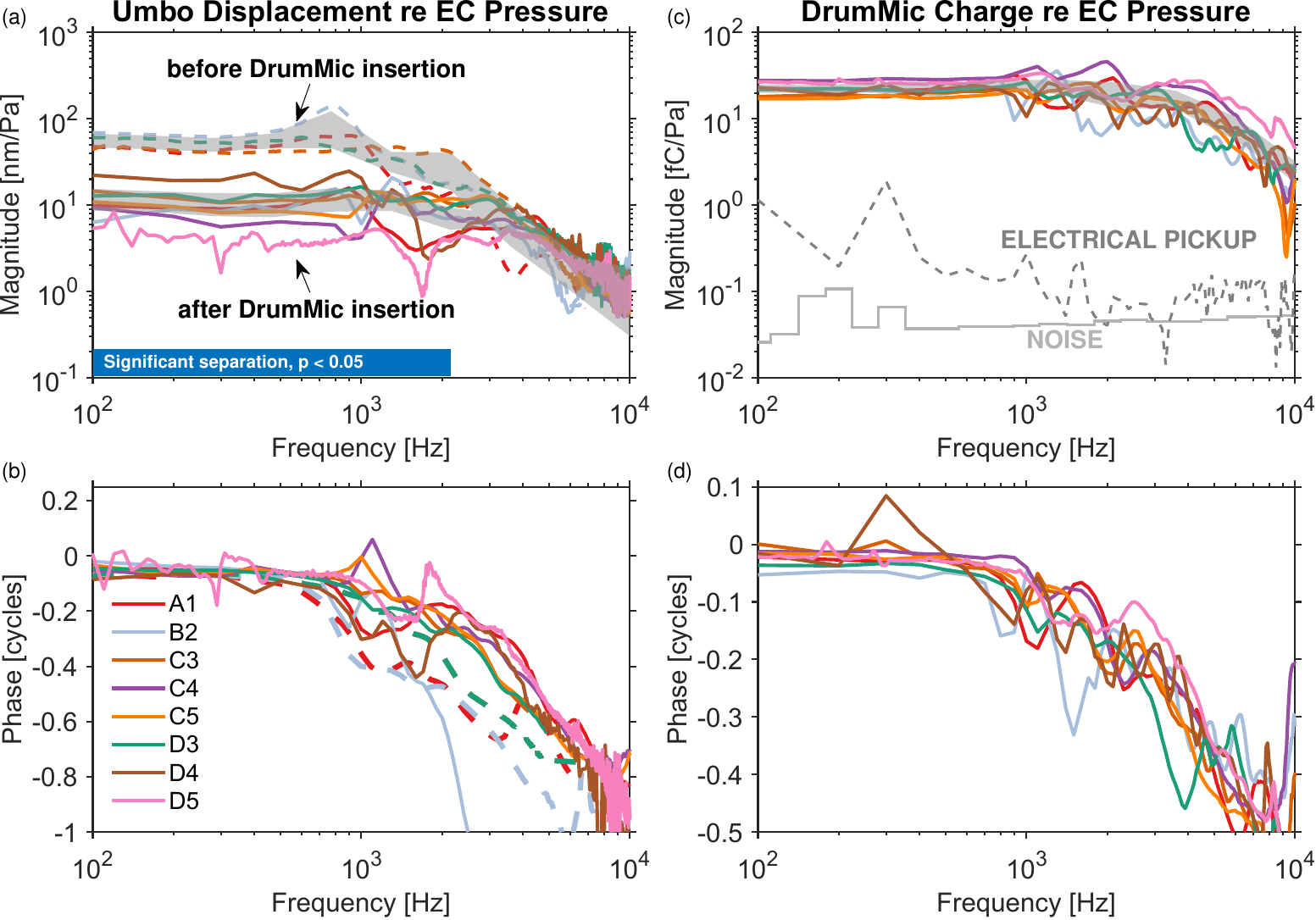}
    \caption{\jz{Umbo displacement (a) magnitude and (b) phase normalized to ear canal \jzz{(EC)} pressure. The blue bar in (a) indicates the frequencies where there is significant separation in umbo displacement between no insertion versus insertion. DrumMic sensitivity (c) magnitude and (d) phase normalized to ear canal pressure. The intrinsic noise of the amplifier and DrumMic is plotted with a solid gray line in (c) and the pickup due to electrical coupling is plotted in a dashed gray line. The shaded gray regions in (a) and (c) show the bootstrapped 95\% confidence interval of the mean for the response magnitudes.}}
    \label{fig:sensitivity}
\end{figure}

\begin{figure}[hbt!]
    \centering
    \includegraphics[width=0.7\linewidth]{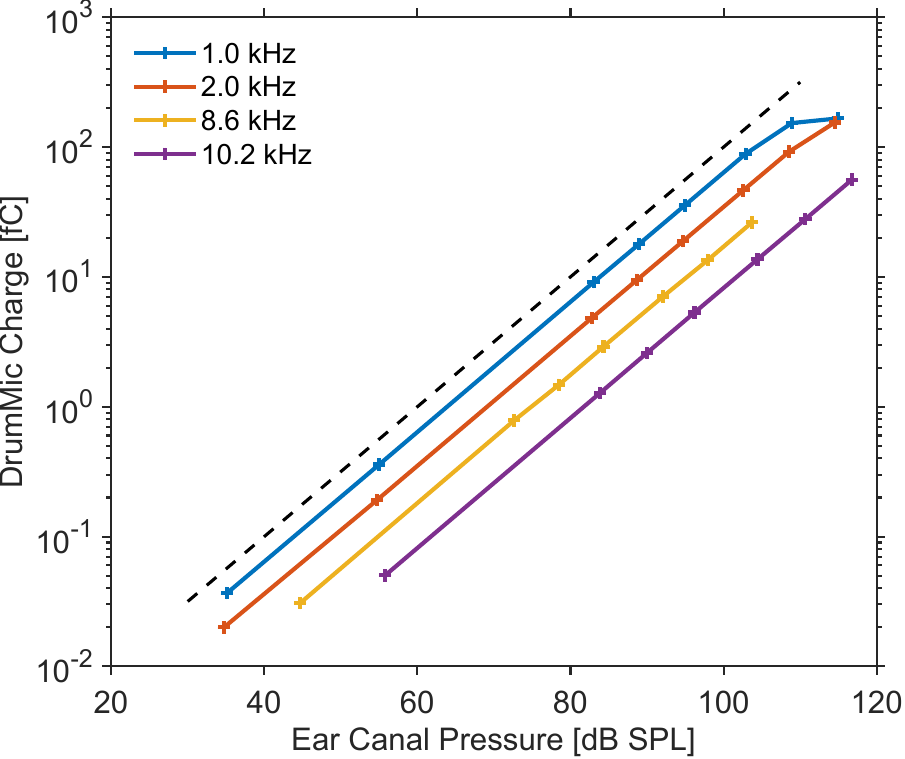}
    \caption{\jzz{DrumMic (D5) input-output curves show linearity up to 100 dB SPL input for different frequencies.} Dashed line is a reference for linearity.}
    \label{fig:linearity}
\end{figure}

\begin{figure}[hbt!]
    \centering
    \includegraphics[width=0.7\linewidth]{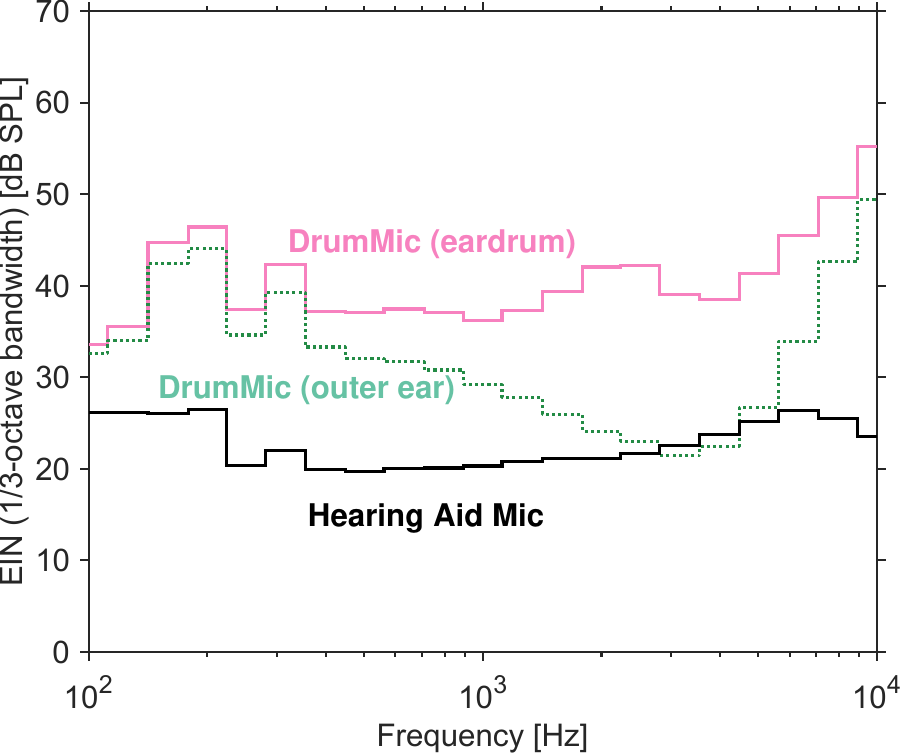}
    \caption{Typical EIN (1/3-octave bandwidth) of an implanted DrumMic (D5) with and without the simulated pressure gain of the outer ear as compared to the an external hearing aid microphone (Knowles EK-23103) that does not leverage the filtering of the outer ear.}
    \label{fig:EIN}
\end{figure}

\jz{Five different microphones (labeled 1,2,3,4,5) were tested in four different ears (labeled A, B, C, D) following the procedure outlined in Methods \Cref{sec:methods}. The results of these experiments are shown in \Cref{fig:offset} through \Cref{fig:EIN}. The key metrics which are reported include sensitivity, linearity, bandwidth, and noise.}

\jz{We studied the effect of drumhead static offset through the use of shims. \Cref{fig:offset} shows the magnitude and phase of the DrumMic charge with respect to umbo displacement (a \& b) or with respect to ear canal pressure (c \& d). In \Cref{fig:offset}, we present the data from experiments D3 and D5. The output of the DrumMic is electrical charge since it is a piezoelectric material connected to a charge amplifier. Note that displacements are reported in nanometers (nm), pressures are reported in pascals (Pa), and the charge is reported in units of femtocoulombs (fC). The sensitivities are normalized quantities and the actual acoustic stimuli used were on the order of \SI{0.3}{Pa} (84 dB SPL) or lower, which were within the the linear range of the DrumMic.}

\jz{\Cref{fig:offset} shows \jzz{a representative experiment}. In this ear, having no shim (gray curves) has the lowest sensitivity with large fluctuations with frequency.  Good contact and some tension between the DrumMic PVDF membrane and umbo was achieved with one shim (green) and two shims (pink). It is evident that the spread in sensitivity with respect to umbo displacement between one and two shims is almost 20 dB (\Cref{fig:offset}a), however, the spread in sensitivity with respect to ear canal pressure has a spread less than 3 dB (\Cref{fig:offset}c).}

\jz{\Cref{fig:sensitivity} presents frequency response data for all experiments across five different DrumMic devices and four different cadaveric ears. \jzz{For achieving a good response, we found that the minimum combined thickness of shims(s) that allowed for contact between the DrumMic and umbo without looseness after DrumMic insertion was sufficient. This resulted in a slight observable static displacement of the umbo causing a static deflection of the DrumMic membrane.} \Cref{fig:sensitivity}a and \Cref{fig:sensitivity}b show the magnitude and phase of the umbo displacement normalized to the input ear canal pressure. The dashed lines indicate the umbo displacement before insertion of the DrumMic and the solid lines indicate response after insertion. At low frequencies there is a 10-20 dB drop in umbo displacement after insertion of the DrumMic. A two-sample t-test shows that there is a significant separation between the umbo displacement before and after insertion up to \SI{2}{kHz}. The ranges (95\% confidence intervals) around the mean  of the umbo motion before and after insertion were both less than 9~dB. Note in \Cref{fig:sensitivity}b, the phase for B2 deviates at higher frequency (possibly due to a complex mode of ossicular motion).}

Plotted in \jz{\Cref{fig:sensitivity}c and \Cref{fig:sensitivity}d, DrumMic sensitivity with respect to ear canal pressure is about \SI{25}{fC/Pa} across five DrumMics and four ears. The frequency response magnitude (\Cref{fig:sensitivity}c) is tightly clustered. The entire range (95\% confidence intervals) around the mean is less than 3 dB. The response magnitude -3 dB point is approximately \SI{5}{kHz}, but the response has good signal-to-noise up to \SI{8}{kHz}. The phase drops by half a cycle after the cut-off frequency.}

\jz{Comparing \Cref{fig:sensitivity}a and \Cref{fig:sensitivity}c it is clear that there was no relationship between umbo motion and DrumMic sensitivity with respect to ear canal pressure for these experiments. The 95\% confidence intervals are smaller for the DrumMic sensitivity (\Cref{fig:sensitivity}c) than for the umbo displacement (\Cref{fig:sensitivity}a). In general, the spread of DrumMic sensitivities is tight and no significant correlation exists between the umbo motion (before or after insertion) and the DrumMic sensitivity. The apparent nature and mechanics of this are explained in Discussion \Cref{sec:discussion}.}

\jz{\Cref{fig:sensitivity}c also displays the intrinsic noise (solid gray line) and the electrical pickup (dashed gray line) of the DrumMic and amplifier. The noise and pickup presented here has units of charge. The noise is reported in 1/3-octave bands. The electrical pickup is the equivalent charge output of the DrumMic in the absence of ear canal pressure (see \Cref{sec:exp_protocol}). Noise and electrical pickup are generally several orders of magnitudes lower than the DrumMic charge output referenced to ear canal pressure (note that the sound stimuli used were less than 84 dB SPL, within the linear range).}

\Cref{fig:linearity} shows a typical example of DrumMic charge output versus ear canal pressure input for several frequencies. Linearity of the DrumMic was maintained throughout a large dynamic range (46 dB SPL to 100 dB SPL). 

\Cref{fig:EIN} compares the EIN (1/3-octave bandwidth) (EIN see \Cref{sec:ein}) of the DrumMic implanted in the middle-ear cavity with the EIN (1/3-octave bandwidth) of a Knowles hearing aid microphone. The EIN (0.1-10 kHz) reported is integrated across a bandwidth of 100 Hz to 10 kHz. Without an outer ear, the best performing implanted DrumMic had a EIN (0.1-10 kHz) of 54 dB SPL. The measured EIN (1/3-octave bandwidth) (pink line in \Cref{fig:EIN}) trends upwards at the higher frequencies and is approximately 20 dB higher than the EIN (1/3-octave bandwidth) of the hearing aid microphone (black line in \Cref{fig:EIN}).

\jzz{However,} the outer ear with pinna and full length of the ear canal would result in a pressure gain to the tympanic membrane from a free-field sound stimulus. If the gain due to the outer ear were taken into consideration, we would expect a \jzz{DrumMic EIN (0.1-10 kHz) of 46 dB SPL (green in \Cref{fig:EIN}). As a comparison, the EIN (0.1-10 kHz) for a representative external hearing aid microphone, such as the Knowles EK-23103, was 34 dB SPL.} Therefore, the DrumMic has the potential of providing performance closer to the current state-of-the-art external microphones used in cochlear implants and hearing aids, with the advantage of being fully implantable.

\section{Discussion}\label{sec:discussion}
\jz{Using a short-term experimental setup in cadaveric temporal bones} we demonstrate that the DrumMic's ring base should be mechanically stable in relation to the cochlear promontory and that the \jzz{PVDF} drum membrane should have good contact to the umbo \jz{to achieve good microphone performance (see \Cref{fig:offset}).} Because the distance between the umbo and promontory varies across ears, the height of the DrumMic system \jz{must vary accordingly}. Additionally, the curvature of the cochlear promontory with respect to the umbo varies across ears. In some ears the promontory is nearly symmetric directly under the umbo with only slight circumferential curvature, while in other ears the promontory is uneven and is asymmetric with respect to the umbo.

When necessary, stabilizing the DrumMic \jzz{base} mechanically on the cochlear promontory with shim(s) and sometimes glue allows the DrumMic to be stable and for the PVDF membrane to be in good contact to the umbo. Mechanically stabilizing the DrumMic results in smoother frequency response of the DrumMic sensitivity. This is likely due to reduction in complex modes of motion, such as rocking, of the DrumMic~\cite{cary2022umbo}.

\jz{DrumMic sensitivity can vary due to static drumhead offset. This variation can affect DrumMic sensitivity in two ways. First, an increased offset distorts the shape of the drumhead due to the umbo pushing against the drum membrane. The steepness of the cone-shaped distortion increases microphone sensitivity as can be seen in the sensitivity relative to umbo displacement plot in \Cref{fig:offset}a. With increased steepness, a subsequent normal displacement of the drumhead by the umbo geometrically leads to increased stretching of the drumhead, and hence increased charge. Second, an increased offset also results in an increased tension applied to the PVDF membrane and umbo. This mechanical loading leads to an increased ossicular stiffness that reduces drumhead motion in response to an acoustic stimulus. The reduction in drumhead motion decreases the apparent sensitivity of the DrumMic \jzz{with respect to ear canal pressure}. In \Cref{fig:offset} and \Cref{fig:sensitivity}, the same DrumMic with a greater static offset must undergo a smaller acoustically-driven displacement for the sensitivity to ear canal pressure to remain the same. In summary, the effect is a cancelling out of these two behaviors resulting in a DrumMic sensitivity \jzz{with respect to ear canal pressure} that works similarly over a large range of offsets. This variation across ears of decreasing umbo motion due to the DrumMic (\Cref{fig:sensitivity}a) has a minor influence on the DrumMic sensitivity with respect to stimulus sound \jzz{measured at the ear canal} (\Cref{fig:sensitivity}c).}

\jz{For an implantable microphone, the critical metric is sensitivity to ear canal pressure. As seen in \Cref{fig:sensitivity}c, the DrumMic sensitivity to ear canal pressure is \jzz{consistent} across multiple DrumMics and different ears for our current fabrication method. Therefore, a major advantage of the DrumMic system is its reliable and repeatable sensitivity with respect to ear canal pressure with low inter-individual variation despite differences in static drumhead offset and \jzz{differences in} umbo motion across ears and DrumMics.}

Our experiments in human cadaveric temporal bones demonstrate the robustness and repeatability of the DrumMic across ears and devices to sense ear canal sound. A bandwidth up to \SI{8}{kHz} with a generally flat region below 3--\SI{5}{kHz} can be achieved. The DrumMics are linear across a large dynamic range from 20 dB SPL to 90 dB SPL across a wide bandwidth.

Part of the success of \jzz{signal-to-noise and consistency of} the DrumMic design is due to its placement at the umbo where the ossicular motion is largest, whereas other middle ear ossicles can have complex modes of motion at reasonably low frequencies. Previous implantable microphone designs that are sensing motion at the head of the malleus or the body of the incus are limited in performance (bandwidth) and are challenging to position consistently and robustly~\cite{chen2004phase,jenkins2007us,hake2021design}.
Complex modes of motion are more problematic further down the ossicular chain and the displacement magnitude is largest at the umbo because of the flexibility of the joints~\cite{ravicz2004mechanisms}. Other middle ear devices face challenges due to these properties of the middle ear, which are avoided by the DrumMic.

\jz{The noise level of the DrumMic system was higher than microphones used in hearing aids (see \Cref{fig:EIN}). The EIN (0.1-10 kHz) was found to be 54 dB SPL for the best DrumMic, which is 20 dB higher than a typical external 
microphone for hearing aids, such as the Knowles EK-23103. The EIN (1/3-octave bandwidth) of the DrumMic trends upwards at the higher frequencies as the DrumMic sensitivity drops off. From the noise analysis of the DrumMic and charge amplifier, we believe the primary increase in intrinsic noise is due to the effect of the moist environment on the sensor impedance values. If the design of the DrumMic included shielding that would prevent the environment from loading the charge amplifier, the EIN would be expected to be greatly reduced. Furthermore, an implanted DrumMic would take advantage of the pinna and full length of the ear canal with the advantageous increase of pressure near the umbo as compared to near the pinna where hearing-aid microphones are situated.}

\jz{A drawback of the DrumMic concept as presented here is the relative uncertainty in mechanical fixation of the sensor due to changing middle ear pressure. The wet environment of the middle ear cavity poses yet another challenge to the DrumMic due to the change in electrical resistance and the necessity of long-term hermetic sealing and biocompatibility. These issues must be addressed for a fully implantable solution.} A further limitation of the prototype design of the DrumMic is that the base of the DrumMic sits on top of the cochlear promontory where Jacobson's nerve (tympanic nerve) courses. This is problematic because it has been reported that this nerve lacks a bony covering in about 80\% of ears~\cite{reddy2019endoscopic,porto1978anatomic,kanzara2014clinical}. \jzz{The Jacobson's nerve provides sensory innervation to the lining of the mesotympanum and Eustachian tube and gives parasympathetic supply to the parotid gland.}

Future design of an implantable microphone that senses umbo motion would require electrical shielding and be able to avoid trauma of the Jacobson's nerve. This can be accomplished through new encapsulation methods and microphone designs. The EIN could be further improved by advances in amplifier design, however, investigation of the sound processing capabilities of cochlear implants may prove that the current suite of implantable microphones we have demonstrated are sufficiently performant. Ongoing progress in implantable microphone research provides hopeful signs that totally implantable cochlear implants may be possible in the near future.

\newpage
\backmatter

\bmhead{Declarations}
The authors declare that they have no conflicts of interest.

\bmhead{Acknowledgments}
This work was supported by the NIH through grant R01DC016874, the NSF through GRFP Grant Number 1745302, the Foundation Prof. Dr. Max Cloëtta, and Research Fund of the University of Basel. We thank Mike Ravicz, Charles Hem, and Anbuselvan Dharmarajan for their help.

\newpage
\bibliography{references}
\end{document}